\documentclass[prl,reprint,amsmath,amssymb,preprintnumbers,superscriptaddress,nofootinbib]{revtex4-1}

\usepackage{epsfig,epstopdf}
\usepackage{subfigure}
\usepackage{color}

\def\beq{\begin{equation}}
\def\eeq{\end{equation}}
\def\bea{\begin{eqnarray}}
\def\eea{\end{eqnarray}}

\newcommand{\lsim}{
\mathrel{\hbox{\rlap{\hbox{\lower4pt\hbox{$\sim$}}}\hbox{$<$}}}}
\newcommand{\gsim}{
\mathrel{\hbox{\rlap{\hbox{\lower4pt\hbox{$\sim$}}}\hbox{$>$}}}}

\newcommand{\dis}[1]{\begin{equation}\begin{split}#1\end{split}\end{equation}}

\begin{document}

\title{de Sitter swampland bound in the Dirac-Born-Infeld inflation model}

\author{Min-Seok Seo}
\email{minseokseo57@gmail.com}
\affiliation{Department of Physics, Chung-Ang University, Seoul 06974, Korea}

\begin{abstract}
\noindent 

 We study the de Sitter (dS) swampland conjecture in the Dirac-Born-Infeld (DBI) inflation model.
 We obtain the dS swampland bound for the relativistic regime using Bousso's entropy bound argument and proper distance.
 It restricts $m_{\rm Pl}\nabla V/V$ by some positive constant depending on warping and the field range. 
 In the specific case of the DBI model driven by the quadratic potential, the model-dependent backreaction argument is interpreted as a natural bound for the slow-roll parameter.
 This shows that quasi-dS spacetime in the DBI model is a result of tuning.

\end{abstract}
\maketitle

\section{Introduction}
 
 Recently, various criteria for the low-energy effective field theory (EFT) admitting a consistent UV completion in quantum gravity have been proposed (for a review, see Ref. \cite{Brennan:2017rbf}).
 Among these ``swampland conjectures," the de Sitter (dS) swampland conjecture \cite{Obied:2018sgi} rules out (meta)stable dS vacua from the string landscape by imposing bounds on the gradient of potential, 
 \dis{&m_{\rm Pl}\frac{|\nabla V|}{V} \geq c,\quad\quad\quad {\rm or}
 \\
&m_{\rm Pl}^2\frac{{\rm min} (\nabla_i\nabla_j V)}{V}\leq -c',\label{Eq:dSconj}}
for some positive constants $c$ and $c'$ of order 1 \cite{Ooguri:2018wrx}.
 As pointed out in Ref. \cite{Ooguri:2018wrx}, the bounds  are supported by  Bousso's covariant entropy bound \cite{Bousso:1999xy}, provided the entropy is dominated by the rapidly increasing number of states along some parameter.
This requirement is guaranteed by another swampland conjecture, the distance conjecture \cite{Ooguri:2006in}.
It states that as a scalar field traverses along the trans-Planckian geodesic distance towers of light states descend from UV.  
 
 Implicit in the arguments of Ref. \cite{Ooguri:2018wrx} is that the dS swampland conjecture imposes constraints on the rate of change of the curvature radius or, equivalently, Hubble parameter $H$. 
  In quasi-dS spacetime, which lasts for a sufficiently large number of $e$-folds, these constraints are written as conjectured bounds for slow-roll parameters,
 \dis{\epsilon_H=-\frac{\dot{H}}{H^2},\quad\quad 
 \tilde{\eta}_H=\frac{\dot{\epsilon}_H}{H\epsilon_H}.\label{Eq:slowH}}
 They measure  deviation of the spacetime geometry from dS, reflecting the instability of dS isometries under dynamics of scalar fields. 
 Such an instability implies that quasi-dS spacetime is a result of fine-tuning, unless there is a specific symmetry reason.
 This fact already appeared as a difficulty in supergravity model building for inflationary cosmology,  namely, the $\eta$ problem \cite{Copeland:1994vg}.
 \footnote{For previous discussions on the dS swampland conjecture in the context of inflationary cosmology, see, e.g., Refs. \cite{Agrawal:2018own, Kehagias:2018uem, Kinney:2018nny, Ashoorioon:2018sqb}.
 For studies on the entropy argument in Ref. \cite{Ooguri:2018wrx}, see, e.g., Ref. \cite{Andriot:2018mav}. }
 
 In Refs. \cite{Obied:2018sgi, Ooguri:2018wrx}, constraints on $(\epsilon_H, \tilde{\eta}_H)$ were stated in terms of the potential as Eq. (\ref{Eq:dSconj}) by assuming the action comprises a quadratic kinetic term and potential only, though the results of Ref. \cite{Ooguri:2018wrx} applies more generally. 
  In this case, slow-roll parameters $(\epsilon_H, \tilde{\eta}_H)$ are related to ``potential slow-roll parameters"
   \dis{\epsilon_V=\frac{m_{\rm Pl}^2}{2}\Big(\frac{V'}{V}\Big)^2,\quad\quad
   \eta_V=m_{\rm Pl}^2\frac{V''}{V},\label{Eq:slowV}}
by $\epsilon_V \simeq \epsilon_H$ and ${\eta}_V \simeq 2\epsilon_H-\tilde{\eta}_H/2$; hence, bounds for $(\epsilon_H, \tilde{\eta}_H)$ are equivalent to those for $(\epsilon_V, \eta_V)$.
 On the contrary, such an equivalence is no longer the case when higher-derivative terms are taken into account.
 The potential may not be an essential ingredient (as in kinematically driven inflation \cite{ArmendarizPicon:1999rj})  or  can be steep (as in Dirac-Born-Infeld (DBI) inflation \cite{Silverstein:2003hf})  to achieve quasi-dS spacetime.
 In these $P(X=-1/2(\partial \phi)^2, \phi)$-type models \cite{Chen:2006nt}, quasi-dS spacetime parametrized by small slow-roll parameters $(\epsilon_H, \tilde{\eta}_H)$ is not a result of small potential slow-roll parameters $(\epsilon_V, \eta_V)$.
 From this, we expect that the dS swampland bound from Ref. \cite{Ooguri:2018wrx} has a nontrivial form, rather than naively given by Eq. (\ref{Eq:dSconj}).

 The purpose of the present work is to explore the results of Ref. \cite{Ooguri:2018wrx} in the context of the DBI inflation model in which a more generic form of the dS swampland conjecture in terms of Eq. (\ref{Eq:slowH}) (implicit in Ref. \cite{Ooguri:2018wrx}) applies.
 In the DBI model, inflation is driven by a modulus of the probe brane moving toward a warped anti-de Sitter (AdS) throat in high speed, so higher-derivative terms in the DBI action play the crucial role.
 Also, masses of open string fluctuations and Kaluza-Klein (KK) modes are proportional to the modulus, realizing the situation similar to what the distance conjecture states.
 This enables us to find the dS swampland bound through the arguments in Ref. \cite{Ooguri:2018wrx}.
 
 In addition, we point out that in the case of quadratic potential, a natural bound for $\epsilon_H$ can be obtained from the IR cutoff for the modulus.
  The IR cutoff here is introduced to protect the EFT from descent of KK modes near the throat.
  This in fact is a part of so-called model-dependent backreaction argument restricting the viable DBI model.
Indeed, it turns out that the original DBI model is easily spoiled by two types of backreaction:
   \begin{itemize}
   \item[1.] Model dependent backreaction \cite{Maldacena} is a deformation of the warp factor by the backreaction from the potential.
   Combined with the IR cutoff, this excludes the DBI model driven by the simple quadratic potential, $V=(1/2) m^2 \phi^2$, unless $m$ is much heavier than the typical mass scale of light particles living on the brane.
   \item[2.] Model-independent backreaction \cite{Chen:2008hz} is a deformation of the warp factor by Hubble expansion of the brane world volume.
   \footnote{In the DBI model, the energy density is dominated by the potential, so deformation of the warped factor by Hubble expansion includes that by the potential.
   Whereas model-dependent backreaction focuses on the viable field range for EFT to exclude the specific quadratic potential, model-independent backreaction considers generic effects of the Hubble parameter.}
    The position of the throat is shifted, and the modulus speed is uncontrollably boosted such that  acceleration stops before a sufficient amount of $e$-folds. 
   \end{itemize}  
 As shown in Ref. \cite{Chen:2008hz}, reduction of the model-independent backreaction effect through nonrenormalizable terms or interaction with bulk fields is regarded as a tuning similar to the attempt to solve the $\eta$ problem.
It is another way to reveal difficulty in quasi-dS spacetime construction. 
 On the other hand, the model-dependent backreaction argument provides a  bound for $\epsilon_H$ parametrizing the tuning for quasi-dS spacetime in terms of the modulus mass.

  Finally, in the nonrelativistic regime the warp factor is interpreted as a potential; hence, we can apply the dS swampland conjecture to it.
  From this, we obtain the upper limit of the dS swampland bound value.
  
 
\section{De Sitter swampland conjecture for Hubble parameter}
  
We begin our discussion with the dS swampland conjecture from Bousso's entropy bound as studied in Ref. \cite{Ooguri:2018wrx}.
 Suppose there are towers of states of which the masses depend on some parameter $\varphi$ as $m=m_0{\rm exp}[-\alpha \varphi]$.
 That means that as $\varphi$ becomes larger than $1/\alpha$ light degrees of freedom descend from UV, invalidating EFT.
From this, we expect that the number of light degrees of freedom increases as $N(\varphi)=n(\varphi){\rm exp}[\beta \varphi]$ for some positive constant $\beta$  to dominate the Hilbert space, hence entropy.
 The behavior $dN/d\varphi>0$ is reflected in the condition for the number of towers $n(\phi)$, $dn/d\varphi \geq 0$, and the exponential factor, which is determined by the details of model.

 On the other hand, the geometry of the Universe close to dS has the curvature radius given by $1/H$.
 Then the entropy can be written as $S=N^p (m_{\rm Pl}/H)^q$ with positive $p$ and $q$ as an ansatz  \cite{Ooguri:2018wrx}.
 When $\varphi \gg 1/\alpha$, the entropy $S$ would eventually saturate its upper limit,  given by the Gibbons-Hawking entropy bound $S_{\rm GH}=8\pi^2 m_{\rm Pl}^2/H^2$. 
   This leads to the generic dS swampland  bound \cite{Ooguri:2018wrx},
 \dis{-\frac{1}{H}\frac{d H}{d \varphi}=\frac{p}{2-q}\frac{1}{N}\frac{dN}{d\varphi}>\frac{p \beta}{2-q}\equiv c_{\rm ent}.\label{Eq:genbound}}
 We note that for $N(\varphi)$ not to be suppressed under the large curvature radius or $H \ll m_{\rm Pl}$, $q < 2$ needs to be satisfied, so the rhs is positive.
 Even if the Universe is dynamical,  it can be approximated as a stable state close to dS, i.e., quasi-dS, provided small $(\epsilon_H, \tilde{\eta}_H)$ is maintained for the sufficiently large number of $e$-folds.
 In this case, the lhs of Eq. (\ref{Eq:genbound}) can be rewritten in terms of the slow-roll parameter $\epsilon_H$ as $-(1/H)dH/d\varphi=\epsilon_H(H/\dot{\varphi})$.
 
 We note that in the argument to obtain Eq. \eqref{Eq:genbound}, an exponential decrease of masses along some parameter is assumed.
 Such a behavior of masses is guaranteed by the distance conjecture.
 Here, the parameter $\varphi$ is identified with the geodesic distance in field space.
 \footnote{We note here the subtlety that,  the physically relevant distance during inflation is the dynamical field range traversed by the inflaton in a multidimendional field space \cite{Landete:2018kqf}, which can be different from the geodesic distance. }
 Typically, exponents $\alpha$ and $\beta$ are given by of order 1 such that the descent of states from UV becomes evident after Planckian excursion of the scalar along the geodesic.
 As $m_{\rm Pl}\varphi$ corresponds to  the variation of the scalar field with the canonical kinetic term, in the absence of  higher-derivative terms in the action, $\varphi$ describes the well-known slow-roll inflation. 
 From relations $m_{\rm Pl}\dot{\varphi}\simeq -V'/3H$ and $\epsilon_H \simeq \epsilon_V$, we obtain $(H/\dot{\varphi})\epsilon_H\simeq (m_{\rm Pl}/2)|V'|/V$, so Eq. (\ref{Eq:genbound}) becomes $m_{\rm Pl}|V'|/V> c_{\rm ent}$, as Eq. (\ref{Eq:dSconj}) states.
 Since the typical value of $c_{\rm ent}$ is of order 1, $\epsilon_V$, and hence $\epsilon_H$, cannot be small enough to maintain quasi-dS spacetime, contradicting our assumption.
 It suggests that inflation is not preferred by quantum gravity.

  In the DBI model we will study, on the other hand, inflaton is the brane modulus, and the masses of open string fluctuations as well as KK modes of particles living on the brane are proportional to it.
  They descend exponentially in terms of  the proper distance in extra-dimensional space, rather than geodesic distance in field space defined by the quadratic kinetic term.
 Therefore, in our discussion, we take the proper distance as the parameter $\varphi$ to apply for the bound, Eq. (\ref{Eq:genbound}).
 \footnote{In the DBI model, the field range of the brane modulus is sub-Planckian, given by $g_{\rm YM}m_{\rm Pl}/\sqrt{N}$, which can be read off from the throat volume contribution to Planck mass \cite{Baumann:2006cd}.
 This is a different situation from what distance conjecture considers, in which the scalar travels the Planckian distance along the geodesic until the descent of light particles.}

\section{DBI inflation Model}
 
 In this section, we review essential features of the DBI inflation model, as studied in Refs. \cite{Silverstein:2003hf, Alishahiha:2004eh}. 
 Let the probe $D3$-brane in type IIB string theory move toward a throat of the warped AdS bulk background generated by a stack of $N$ $D3$-branes.
 Given the distance of the probe brane from the throat  $r \equiv \alpha' \phi$, where $\alpha'$ is the Regge slope   representing (string length)$^2$, the metric is given by
 \dis{\frac{ds^2}{\alpha'}=f^{-1/2}(\phi)[-dt^2+a(t)^2d\vec{x}^2]+f^{1/2}(\phi)[d\phi^2+\phi^2 d\Omega_5^2].\label{Eq:AdS}}
 Here, $f(\phi)=2\lambda/\phi^4$ is the warp factor, with $\lambda=N g_{\rm YM}^2$ being the 't Hooft coupling for the gauge interaction on the brane.
 Assuming spatial homogeneity, the effective action for $\phi$ is written as
 \dis{S=-\frac{1}{g_{\rm YM}^2}\int d^4x \sqrt{-g}\Big[f^{-1}(\phi)[\sqrt{1-f(\phi)\dot{\phi}^2}\mp 1]+V(\phi)\Big].\label{Eq:DBIaction}}
 The first term comes from the DBI action, which contains  the gravitational interaction.
   From this, one finds that the speed of $\phi$ cannot be arbitrarily large but restricted to be $f\dot{\phi}^2\leq 1$ \cite{Kabat:1999yq}.
 The second term, the Chern-Simons term, describes the exchange of the four form Ramond-Ramond (RR) sector field between branes, where the upper (lower) minus (plus) sign is assigned for the probe (anti-)$D3$-brane.
 While a gauge symmetry in this case is given by $U(1)\times U(N)$, when the probe brane reaches the AdS throat, it is enhanced to $U(N+1)$.
 Then, the Higgsed gauge bosons as well as matters which are bifundamental under $U(1)\times U(N)$ become massless as their masses are given by $\phi$.
 Higher-derivative terms are the effect of these  particles in the virtual loop.

 From Eq. (\ref{Eq:AdS}), we define the dimensionless proper distance in the $\phi$ direction as $ds=-(2{\alpha'}^2\lambda)^{1/4}m_{\rm Pl}d\phi/\phi$, or
 \dis{\phi=\phi_0{\rm exp}\Big[-\frac{s}{(2{\alpha'}^2\lambda)^{1/4} m_{\rm Pl}}\Big].\label{Eq:distance}}
 Here, the negative sign indicates that $\phi$ gets smaller as the probe brane approaches the AdS throat.
 That means that, before reaching the AdS throat ($\phi=0$), $\phi$ travels along infinitely long proper distance $s/m_{\rm Pl}$, and when it exceeds the AdS radius $(2{\alpha'}^2\lambda)^{1/4}$, masses of gauge bosons and bifundamentals become small enough.
 At the same time, as we will see, the tower of KK modes also descends from the UV.
 This suggests regarding the proper distance $s$ as the parameter $\varphi$ in Eq. (\ref{Eq:genbound}).

 To see the effects from higher derivatives of $\phi$ in detail, we define the boost factor in a way similar to that in special relativity,
 \dis{\gamma=\frac{1}{\sqrt{1-f(\phi)\dot{\phi}^2}}.}
 The relativistic limit corresponds to $\gamma \gg 1$.
  The equations of motion for the probe brane are given by \cite{Silverstein:2003hf}, \footnote{For the probe antibrane, $V$ is replaced by by $V+2f^{-1}$.}
 \dis{&3 H^2=\frac{\rho}{g_{\rm YM}^2 m_{\rm Pl}^2},\quad\quad 2\frac{\ddot{a}}{a}+H^2=-\frac{p}{g_{\rm YM}^2 m_{\rm Pl}^2},
 \\
 &\rho=f^{-1}(\gamma-1)+V,\quad\quad p=\frac{\gamma-1}{\gamma}f^{-1}-V,
 \\
 &\ddot{\phi}+\frac{3f'}{2f}\dot{\phi}^2-\frac{f'}{f^2}+\frac{3H}{\gamma^2}\dot{\phi}+\Big(V'+\frac{f'}{f^2}\Big)\frac{1}{\gamma^3}=0.\label{Eq:DBIeom}}
  From them, one finds the condition for the Universe to accelerate:
\dis{2\frac{\ddot{a}}{a}=\frac{1}{g_{\rm YM}^2 m_{\rm Pl}^2}\Big[\frac23V-\frac{(\gamma-1)(\gamma+3)}{3\gamma}f^{-1}\Big]>0.}
 Therefore, even in the relativistic regime $\gamma \gg 1$, the scale factor $a(t)$ accelerates, provided the ``large $V$ parameter" defined by
 \dis{c_V \equiv \frac{\gamma}{f V}  \label{Eq:Infcond},}
 is much smaller than unity,  such that the potential gives the dominant contribution to the energy density.
 
\section{DS swampland conjecture in DBI - relativistic regime }

 Now, we observe the dS swampland conjecture in the relativistic regime.
 To see this, we first find how $\epsilon_H$ is written in terms of the large $V$ parameter $c_V$.
 From time derivatives of the first in Eq. (\ref{Eq:DBIeom}) and $\gamma$, with the help of the last in Eq. (\ref{Eq:DBIeom}), we obtain $\dot{\phi}^2=-2 g_{\rm YM}^2 m_{\rm Pl}^2 \dot{H}/\gamma$ \cite{Silverstein:2003hf}.
 Then, $\epsilon_H$ in the regime $\gamma \gg 1$ is given by
 \dis{\epsilon_H &=-\frac{\dot{H}}{H^2}=\frac32\frac{(\gamma^2-1)/\gamma}{f V+\gamma-1} \simeq \frac32\frac{c_V}{1+c_V}. \label{Eq:epsilon}}
 Evidently, $c_V=0$ corresponds to dS spacetime. 
For $c_V \ll 1$, the Universe accelerates, and the probe brane has a quasi-dS geometry satisfying $\epsilon_H \simeq (3/2)c_V \ll 1$. 
Moreover, the value of $\epsilon_H$ converges to $3/2$ for large $c_V$, so the dS swampland condition in the relativistic regime is written as $\epsilon_H \sim {\cal O}(1)$, rather than $\epsilon_H > {\cal O}(1)$ as naively expected from Eq. (\ref{Eq:dSconj}).
 The consideration above suggests that the large $V$ parameter $c_V$ is a good slow-roll parameter measuring the deviation of spacetime from dS, which is equivalent to $\epsilon_H$ in quasi-dS.
 Then, the dS swampland conjecture for the DBI model in the relativistic regime can be given by the bound for $c_V$.

 We also note that when we trade the time dependence of $H$ into $\phi$ dependence, $H(\phi)$, $\dot{H}=H'\dot{\phi}$ gives $\dot{\phi}=-2 g_{\rm YM}^2 m_{\rm Pl}^2 H'/\gamma$, from which we obtain $\gamma^2=1+f(2 g_{\rm YM}^2 m_{\rm Pl}^2 H')^2$.
Since the energy density is dominated by $V$, the first equation of motion gives $6 g_{\rm YM}^2 m_{\rm Pl}^2 HH'\simeq V'$, and then the relativistic condition reads \cite{Silverstein:2003hf}
\dis{\gamma^2\simeq 1+f g_{\rm YM}^2 m_{\rm Pl}^2\frac{(V')^2}{3V} \simeq \frac23 \epsilon_V g_{\rm YM}^2 f V \gg 1.}
That means that during the quasi-dS evolution, the potential is rather steep, contrary to the situation in the absence of higher-derivative terms in which quasi-dS is a result of the almost flat potential.
 Combining this with  Eq. (\ref{Eq:Infcond}), the definition of $c_V$, we obtain the relation for $\gamma \gg 1$ \cite{Alishahiha:2004eh},
  \dis{c_V \gamma \simeq \frac43 g_{\rm YM}^2 m_{\rm Pl}^2\Big(\frac{H'}{H}\Big)^2 \simeq \frac23 g_{\rm YM}^2 \epsilon_V.\label{Eq:epgam}}
  This shows that  large $\gamma \gg 1$ and perturbatively small $g_{\rm YM}^2 <4\pi$ allow for $c_V \ll 1$, i.e., quasi-dS unless   $\epsilon_V$ is much larger than $\gamma/g_{\rm YM}^2 \gg 1$.

From the discussion so far, we find how the generic dS swampland bound in Eq. (\ref{Eq:genbound}) appears in the DBI model.
Putting the first two terms in Eq. (\ref{Eq:epgam}) as well as the definition of the proper distance $s'=-(2{\alpha'}^2\lambda)^{1/4}m_{\rm Pl}/\phi$ into $dH/ds=H'/s'$ in the lhs  of Eq. (\ref{Eq:genbound}), we obtain the dS swampland bound
\dis{(c_V\gamma)^{1/2}>(m_{\rm Pl}(2{\alpha'}^2\lambda)^{1/4})\Big(\frac{g_{\rm YM}m_{\rm Pl}}{\phi}\Big)\frac{2}{\sqrt3}c_{\rm ent}. \label{Eq:DBIdS0}}  
The first term in the rhs corresponds to the characteristic constant for the proper distance, the AdS radius.
 As for $g_{\rm YM} m_{\rm Pl}/\phi$, the throat volume contribution to $m_{\rm Pl}$ sets the bound $g_{\rm YM} m_{\rm Pl}/\phi>\sqrt{N}$ \cite{Baumann:2006cd}.
 Therefore, $(c_V\gamma)^{1/2}$ or, equivalently, $g_{\rm YM}m_{\rm Pl}\nabla V/V$ [from the first and third terms in Eq. (\ref{Eq:epgam})]  is bounded  by the AdS radius and $g_{\rm YM} m_{\rm Pl}/\phi$ bound in addition to $c_{\rm ent}$, rather than simply bounded by $c_{\rm ent}$ only.
 Whether quasi-dS is allowed, i.e., $c_V$ can be still smaller than unity under the dS swampland bound, is determined by the tuning between the rhs of Eq. (\ref{Eq:DBIdS0}) and $\gamma$.

  We close this section with comments on the condition that $\tilde{\eta}_H$  remains small. 
  Taking the time derivative on Eq. (\ref{Eq:epgam}) gives the expression for $\tilde{\eta}_H$ under $\gamma \gg 1$ and $c_V\ll 1$ as
  \dis{\tilde{\eta}_H&=\frac{2g_{\rm YM}^2 m_{\rm Pl}}{\gamma}\Big[2\frac{{H'}^2}{H^2}-2\frac{H''}{H}+\frac{\gamma'}{\gamma}\frac{H'}{H}\Big]
  \\
  &=3\epsilon_H-\frac{g_{\rm YM}^2m_{\rm Pl}^2}{\gamma}\frac{V''}{V}-\frac{2g_{\rm YM}^2 m_{\rm Pl}^2}{\gamma}\frac{V'}{\phi V},}
  where for the last equality $\phi$ derivatives on $3g_{\rm YM}^2 m_{\rm Pl}^2 H^2 \simeq V$ and $\gamma^2=(g_{\rm YM}^2m_{\rm Pl}^2/3)f({V'}^2/V)$ are taken.
  The last term, which can be rewritten as $\sim (g_{\rm YM} m_{\rm Pl}/\phi)(\epsilon_H/\gamma)^{1/2}$, can be made small through tuning between $g_{\rm YM}m_{\rm Pl}/\phi$ (bounded by $\sqrt{N}$) and  $(\epsilon_H/\gamma)^{1/2}$ (much smaller than 1).
 Also,  in the second term, sizeable $m_{\rm Pl}^2 \nabla^2 V/V$ is allowed in quasi-dS spacetime as long as it is much smaller than $\gamma/g_{\rm YM}^2$.
  Indeed, for the potential in the form of $V \sim \phi^p$, we simply obtain $\tilde{\eta}_H =[p(p-2)/p^2]\epsilon_H$.
  Then, for the constant and quadratic potential, $\tilde{\eta}_H$ vanishes, and for other values of $p$, the bound for $\tilde{\eta}_H$ comes from that for $\epsilon_H$.

\section{ $c_V$ bound for quadratic potential}

 Since various particles become massless as the probe brane approaches the throat, EFT for the DBI inflation is no longer valid near the throat.
 Moreover, the warp factor can be deformed by the backreaction of large $V$ on the background.
 Juan Maldacena argued that, given quadratic potential $V=(1/2) m^2\phi^2$, for values of $\phi$ which do not alter the warp factor, the large $V$ condition $c_V \ll 1$ is easily spoiled and hence quasi-dS spacetime is difficult to realize \cite{Maldacena} (see Ref.  \cite{McAllister:2007bg} for a review of the argument).
As will be clear, his argument sets the bound for $c_V$ in another way, revealing the nature of the quasi-dS spacetime as a tuning. 
 This can be regarded as another version of the dS swampland bound, limited to the DBI model driven by the quadratic potential.
 
To begin with, suppose some of the massless string excitations on the brane obtain masses, say, $m_0$ through e.g., the supersymmetry breaking.
Now, the KK mass gap for the particles living on the brane at $r=\alpha'\phi$ is given by $m_{\rm KK}=(1/(\alpha'\sqrt{2\lambda})^{1/2})\times (r/(\alpha'\sqrt{2\lambda})^{1/2})$, i.e., the inverse of the AdS radius multiplied by the warp factor.
 Hence, as pointed out in Ref. \cite{Maldacena}, when $r$ gets close to zero, so does $m_{\rm KK}$, resulting in degeneracy of all KK modes of masses $m_n^2=m_0^2+n^2 m_{\rm KK}^2$ ($n \in \mathbb{Z}$) to $m_0^2$.
 Even  worse, KK modes of massless particles like (unbroken) gauge bosons degenerate to zero mass. 
 
 For this reason, EFT for particles with masses around and below $m_0$ are intact by such overcrowding of KK modes, provided $m_0 <m_{\rm KK}(r=r_{\rm IR})$.
 Then, the range of $\phi$ is restricted as  \cite{Maldacena}
   \dis{\phi^2>\phi_{\rm IR}^2=\frac{r_{\rm IR}^2}{{\alpha'}^2}> m_0^2 2\lambda,}   
 from which Ref. \cite{Maldacena}  found a bound for $c_V$,
  \dis{c_V=\frac{\gamma}{fV}=\gamma \frac{\phi^4}{2\lambda}\frac{2}{m^2\phi^2}=\frac{\gamma \phi^2}{\lambda m^2}>2\gamma\Big(\frac{m_0}{m}\Big)^2.\label{Eq:modelbound}}
 Therefore, unless the modulus mass $m$ is considerably heavier than $m_0$, we have $V<\gamma f^{-1}$, inconsistent with the large $V$ condition.
     While the $m\to \infty$ limit corresponds to dS spacetime, $m$ cannot be arbitrarily large as the DBI model is based on EFT which is valid below the mass scale $\phi$.
  In addition, $\gamma \gg 1$ also prevents $c_V$ from being small enough.   
  If $m\simeq m_0$, the slow-roll parameter $c_V$ is forced to be larger than $2\gamma \gg 1$, and quasi-dS spacetime is not allowed. 
   Comparing Eq. (\ref{Eq:modelbound}) with Eq. (\ref{Eq:epgam}), we find that the $c_V$ bound  is converted into the bound for $m_{\rm Pl}\nabla V/V$ as
  \dis{m_{\rm Pl}\frac{V'}{V}=\sqrt{2\epsilon_V}>  \frac{\sqrt6\gamma}{g_{\rm YM}}\Big(\frac{m_0}{m}\Big).}
  
  To explain the natural hierarchy between $m$ and $m_0$, Ref. \cite{McAllister:2007bg} suggested that these two masses are generated in different ways.
  For example, $m$ and $m_0$ may come from gauge and gravity mediation, respectively.
  In any case, the $c_V$ bound from the model-dependent backreaction argument shows that quasi-dS spacetime is a result of tuning between small $m_0/m$ and large $\gamma$.

\section{DS swampland conjecture in DBI - nonrelativistic regime }

  In the nonrelativistic regime ($f(\phi)\dot{\phi}^2 \ll 1$), since higher-derivative terms are suppressed,  the action  given by Eq. (\ref{Eq:DBIaction}) is expanded as 
 \dis{S=\frac{1}{g_{\rm YM}^2}\int d^4 x \sqrt{-g}\Big[\frac12\dot{\phi}^2-f^{-1}\pm f^{-1}-V+\cdots\Big].\label{Eq:DBINR}}
 To find the dS swampland bound from Eq.  (\ref{Eq:genbound}), we consider $-(1/H)dH/ds=\epsilon_H(H/\dot{s})$.
 We note from Eq. (\ref{Eq:DBINR}) that the canonically normalized modulus field is $\phi/g_{\rm YM}$, so Planck mass $m_{\rm Pl}$ in equations of motion for $\phi$ appears as a combination $g_{\rm YM}m_{\rm Pl}$, as can already be found in Eq. (\ref{Eq:DBIeom}).
 Also, in the nonrelativistic limit, the potential slow-roll parameter $\epsilon_V$ is related to $\epsilon_H$ by $\epsilon_H=g_{\rm YM}^2\epsilon_V$.
 Using the equation of motion $\dot{\phi}\simeq -V'/3H$, the dS swampland bound in the nonrelativistic regime is written as
 \dis{\epsilon_H^{1/2}= \epsilon_V^{1/2}> \sqrt2(m_{\rm Pl}(2{\alpha'}^2\lambda)^{1/4})\Big(\frac{g_{\rm YM}m_{\rm Pl}}{\phi}\Big)c_{\rm ent}.\label{Eq:dSDBINR}}
 This, in fact, is the same as the $\gamma\simeq 1$ limit of Eq. (\ref{Eq:DBIdS0}).

 On the other hand, Eq. (\ref{Eq:DBINR}) shows the exact cancellation between gravitational and RR four-form potentials $\mp f^{-1}$ in the probe brane action, reflecting the Bogomolny-Parasad-Sommerfeld nature of the D-brane.
 Hence, at least in the nonrelativistic regime, the dS swampland conjecture  can be applied to the ``potential" $f^{-1}$.
 Then, Eq. (\ref{Eq:dSDBINR}) becomes the condition $c_{\rm ent}<2/[m_{\rm Pl}(2{\alpha'}^2\lambda)^{1/4}]$.
 
 As suggested in Ref. \cite{Agrawal:2018own}, one may try to impose Eq. (\ref{Eq:dSconj}) on $f^{-1}$ as a condition for $\phi$ even though the entropy bound argument with the proper distance does not support it.  
 This conjectured bound  is written as
 \dis{&g_{\rm YM}m_{\rm Pl}\frac{|(f^{-1})'|}{f^{-1}}=\frac{4g_{\rm YM}m_{\rm Pl}}{\phi}>c\label{Eq:WGC}}
 We recall that $\phi$ is interpreted as the mass of $U(1)\times U(N)$ bifundamentals.
 Especially, since the bifundamentals are charged under the unbroken $U(1)$ gauge symmetry, the inequality is translated into  $g_{\rm YM}> c \phi/m_{\rm Pl}= c\times (U(1)~{\rm charged~particle~mass})/m_{\rm Pl}$.
For $c\sim {\cal O}(1)$, the inequality in Eq. (\ref{Eq:WGC}) is more or less consistent with the weak gravity conjecture (WGC) \cite{ArkaniHamed:2006dz} in which the bound is given by the extremal black hole charge-to-mass ratio of order 1.
Of course, this WGC-like bound is not as stringent as the bound obtained in Ref. \cite{Baumann:2006cd}, which is given by  $g_{\rm YM} m_{\rm Pl}/\phi >\sqrt{N}$, since almost static AdS background requires  large $N$. 

 The similarity between bounds in Eq. (\ref{Eq:WGC}) and the WGC is already found in another inflation model.
  Indeed, the WGC has been used to constrain the slow-roll in natural inflation \cite{Freese:1990rb} that makes use of the axionlike pseudo-Goldstone boson as an inflaton.
  Given axion decay constant $f_a$ with potential $V=V_0 {\rm exp}[-S_{\rm int}]\cos(a/f_a)+$(suppressed higher harmonics), we have $m_{\rm Pl}^2 V''/V\sim -m_{\rm Pl}^2/f_a^2$ as well as $m_{\rm Pl}(|V'|/V)\sim m_{\rm Pl}/f_a$, so $f_a$ needs to be trans-Planckian  to give quasi-dS spacetime, but the (refined) dS swampland conjecture with $c, c'\sim {\cal O}(1)$ forbids it.
  \footnote{There is an argument that the distance conjecture applies to the axionlike particles through the saxion backreaction \cite{Baume:2016psm}.
   Then the dS swampland conjecture for the axionlike particles is supported by Ref. \cite{Ooguri:2018wrx}.}
  At the same time, in the WGC, $1/f_a$ and $S_{\rm inst}$ are interpreted as charge and mass respectively, giving a bound $f_a\times S_{\rm inst} \lesssim m_{\rm Pl}$ \cite{Brown:2015iha}.
  In order that terms containing higher harmonics are sufficiently suppressed compared to the leading term of the potential, we need $S_{\rm inst} \sim {\cal O}(1)$, which excludes the single field natural inflation 
  (for the WGC in the context of multifield natural inflation, see, e.g., Refs. \cite{Brown:2015iha, Brown:2015lia}).
  Even though the origins are different, the similarity between two conjectures in two inflationary scenarios in the nonrelativistic regime may imply an equivalence argument  based on quantum gravity.

\section{Summary}

Discussions so far show that in the DBI model Bousso's entropy bound argument provides the bound for $(c_V\gamma)^{1/2}/g_{\rm YM}\simeq  m_{\rm Pl}\nabla V/V$ as Eq. (\ref{Eq:DBIdS0}) but it does not coincide with the $m_{\rm Pl}\nabla V/V$ bound in the nonrelativistic regime.
On the other hand, the model-dependent backreaction argument is interpreted as the $c_V$ bound in the case of quadratic potential. 
From this,  we find that quasi-dS spacetime is a result of fine-tuning between the small mass ratio $m_0/m$ and the large boost factor $\gamma$.

 We note that in the DBI model, the energy density is dominated by the potential, resulting in a connection between the dS swampland condition (Eq. (\ref{Eq:DBIdS0})) and $m_{\rm Pl}\nabla V/V$ bound.
 On the other hand, in the kinematically driven inflation model \cite{ArmendarizPicon:1999rj}, the potential is not required to realize quasi-dS spacetime.
 The dS swampland bound in this case is expected to be stated irrelevant to the $m_{\rm Pl}\nabla V/V$ bound.
 This implies that we may have a more generic bound for the $P(X, \phi)$-type model, but it is still challenging to find out the parameter along which the mass decreases exponentially in a sensible field range if it is sub-Planckian.

\vspace{5mm}
\begin{acknowledgments}

Acknowledgments: 
M.S. is grateful to Gary Shiu for insightful discussion and comments and Jaewon Song for reviewing the draft.
M.S. is supported by NRF Basic Science Research
Program (Grant No. NRF-2016R1A2B4008759) and CAU Research
Grants in 2018.

\end{acknowledgments}


\end{document}